\pdfoutput=1

\documentclass[11pt]{article}

\usepackage{acl}

\usepackage{times}
\usepackage{latexsym}
\usepackage[T1]{fontenc} 
\usepackage[utf8]{inputenc} 
\usepackage{microtype} 
\usepackage{hyperref}
\usepackage{xcolor}
\usepackage{tabularx}
\usepackage{multirow}

\title{NanoNER: Named Entity Recognition for nanobiology using experts' knowledge and distant supervision}

\author{Ran Cheng, Martin Lentschat, Cyril Labbé \\
    Univ. Grenoble Alpes, CNRS, Grenoble INP, LIG, 38000 Grenoble, France \\
    \texttt{\{martin.lentschat,cyril.labbe\}@univ-grenoble-alpes.fr}
     }

\begin{document}

\maketitle

\begin{abstract}
Here we present the training and evaluation of NanoNER, a Named Entity Recognition (NER) model for Nanobiology.
NER consists in the identification of specific entities in spans of unstructured texts and is often a primary task in Natural Language Processing (NLP) and Information Extraction.
The aim of our model is to recognise entities previously identified by domain experts as constituting the essential knowledge of the domain.
Relying on ontologies, which provide us with a domain vocabulary and taxonomy, we implemented an iterative process enabling experts to determine the entities relevant to the domain at hand.
We then delve into the potential of distant supervision learning in NER, supporting how this method can increase the quantity of annotated data with minimal additional manpower.
On our full corpus of 728 full-text nanobiology articles, containing more than 120k entity occurrences, NanoNER obtained a F1-score of 0.98 on the recognition of previously known entities.
Our model also demonstrated its ability to discover new entities in the text, with precision scores ranging from 0.77 to 0.81.
Ablation experiments further confirmed this and allowed us to assess the dependency of our approach on the external resources.
It highlighted the dependency of the approach to the resource, while also confirming its ability to rediscover up to 30\% of the ablated terms.
This paper details the methodology employed, experimental design, and key findings, providing valuable insights and directions for future related researches on NER in specialized domain. 
Furthermore, since our approach require minimal man-power, we believe that it can be generalized to other specialized fields.
\end{abstract}

\section{Introduction}


As the volume of the scientific literature increases, the demand for NLP models able to deal with domain vocabulary and specific knowledge is becoming increasingly apparent.
The NanoBubbles \footnote{\url{https://nanobubbles.hypotheses.org/}} project, from which the work presented here originates, aims at studying \textit{how, when and why science fails to correct itself}.
It focuses on the nanobiology domain and combines approaches from the natural sciences, natural language processing and social sciences.
The field of nanobiology being characterized by both its multidisciplinarity and its high degree of specialization is a perfect example of the need for specialized tools.
Thus, we must leverage methods from Natural Language Processing (NLP) to assist in the extraction of important information from a large number of articles.
The main task of this paper is to train a Named Entity Recognition (NER) model in the field of nanobiology.

The primary task of Named Entity Recognition is to identify and classify specific entities (i.e. named entities) in a text.
Compared to other fields, Biomedical NER (BMNER) is a particularly challenging problem, mainly due to the high cost of obtaining quality annotated data and the complexity of domain terminology.
A famous example of a model able to perform BMNER is bioBERT \citep{lee2020biobert}, which is pre-trained on a large-scale corpus of biomedical text.
It performs well on a standard set of biomedical benchmarks in several downstream tasks (e.g., NER, Relations Extraction, Q\&A).
To our knowledge, NER in the nanobiology domain remains an uncharted territory, as existing BMNER models are not trained to recognize entities of interest in this specific field.

Training an efficient NER model requires a large amount of annotated data, which is not easy to come by in specialized domains as the manual work it requires need to be carried out by fields experts.
In our work for NER in the nanobiology field, we use distant supervision to alleviate for the lack of annotated data and thus allow the creation of a corpus of articles from a specialized domain large enough to train a NER model.
Using BioBERT \cite{lee2020biobert} as base model, this approach requires minimal human work.
We believe that the approach we implemented, and describe here, is adapted to other scientific domain.

First, we harnessed existing nanobiology ontologies (i.e., the Nanoparticle ontology \cite{NPO} and eNanoMapper \cite{hastings2015enanomapper}) for their concept hierarchy and vocabulary.
Then, an iterative process took place with a team of domain experts, who determined the essential labels for our NER model and curated the vocabulary.
A round of vocabulary extension, with expert curation, took place before the automatic annotation of the corpus.
Ablation experiments were also implemented to measure the influence of the vocabulary coverage in our distant supervision setting.

In summary, the main contributions of this paper are as follows:
\begin{enumerate}
    \item We have implemented a method to create annotated data for NER.
    It consists in an iterative process, involving ontology and corpus analysis followed by use of expert knowledge and their validation.
    This lead us to identify five labels, with vocabularies covering 1438 terms, that are highly relevant to nanobiology.
    \item We created NanoNER, a NER model for nanobiology using a distant supervision learning approach and trained on automatically annotated entities in a corpus of 728 unlabelled full-text nanobiology articles.
    Detailed ablation experiments were conducted to evaluate the influence of the vocabulary coverage.
    \item Finally, ablation experiments allowed us to estimate the dependency of our model to the annotation resource.
    We can effectively measure how well NanoNER is capable to generalize, i.e. its ability to (re)find entities not present in the training set, as well as the essential and minimal terms needed to obtain satisfactory results.
\end{enumerate}

\section{Related work}

Existing BMNER solutions encompass early NER methods, such as dictionary matching or rule-based approaches, as well as supervised machine learning methods such as Markov models \citep{ponomareva2007biomedical}.
Conditional Random Fields (CRFs) were then employed to perform BMNER \citep{ponomareva2007conditional,friedrich2006biomedical}. 
Unlike Markov models, CRFs can consider the characteristics of the entire input sequence, not just the current state.
And Support Vector Machine (SVM) can be used in binary classification problems for NER tasks, such as determining whether a word is a named entity of a particular type \citep{ju2011named}.

Recently, deep learning approaches using large amounts of labeled data, such as models built on BioBERT \citep{lee2020biobert}, have achieved state-of-the-art results on BMNER.
For instance, on the jnlpba \cite{Huang_2020} dataset, the KeBioLM model \citep{yuan2021improving} obtained a F1 score of 0.82 on recognizing entities relating to proteins, genes and cells.
In the bc5cdr \cite{bc5cdr} dataset, the BINDER \citep{zhang2022optimizing} model using a contrastive learning approach, achieved a F1 score of 0.91 on chemical and disease entities.
However, BMNER presents specific difficulties.
For instance, \citet{7747810} conducted an extensive study on electronic medical records and identified that such technical texts often contain a substantial amount of specialized terminology and knowledge, and frequently present issues such as spelling errors, abbreviations, and idiosyncratic terms, all of which add to the difficulty of the NER task.
In this difficult setting, they proposed a method based on CNN (Convolutional Neural Networks) and Word2Vec for performing BMNER and managed to achieve a F1 score of 0.73.

To address the scarcity of annotated data in deep learning models, some weak supervision and distant supervision solutions have been proposed.
\citet{mintz2009distant} were among the pioneers of distant supervision learning, introducing this method in information and relation extraction tasks.
Their goal was to extract relations between entities from a large amount of unlabeled text, using existing knowledge bases as distant supervision signals.
Distant supervision was initially widely applied to relation extraction tasks and later extensively used in NER tasks.
Distant supervision methods for NER have been validated in previous studies.
\citet{shang2018learning} revised the LSTM-CRF NER model of \citet{lample2016neural} and utilized the MeSH database for chemical and disease entity research.
Since the automatic annotation of a corpus tend to introduce noise in the training data, some methods have been proposed to reduce this effect \citep{meng2021distantly}, such as using early stopping or introducing the concept of pseudo-labels \citep{liang2020bond}.
Early stopping prevent overfitting the model on the training data and fosters the learning of important features of the corpus.
Pseudo-labeling data expand the training set by generating new labeled data that can then be used alongside existing datasets.

BMNER using ontologies and distant supervision have already been performed in the biomedical domain \cite{fries2017swellshark,wang2021chemner} and this type of approach could be generalized to any domain for which a semantic and lexical resource exists.
These works used different technics to minimize the risk of noise propagation, e.g. filtering candidate annotations through heuristics based on part-of-speech analysis \cite{fries2017swellshark} or disambiguating ambiguous entities based on other entities present in the same context \cite{wang2021chemner}.
In our work, we rely on domain experts at crucial steps: (1) determining the labels and then (2) filtering and validating the vocabulary of our annotation resource.

To the best of our knowledge, no one has yet proposed a BMNER model that meets the information mining needs of the nanobiology field.
We thus aim at training a NER model, using minimal man-power, but which still meets experts requirements regarding the entities of interest of the domain.

\section{Data preparation}



Here we describe the essential resources for our work, the corpus and ontologies used, the expert work on selecting labels relevant to the domain at hand as well as the vocabulary associated, and the automatic annotation on the scientific articles.
All codes necessary to replicate this study are available online\footnote{\url{https://gricad-gitlab.univ-grenoble-alpes.fr/nanobubbles/nano-ner-wiesp-2023.git}}.

\subsection{Corpus}

The corpus used in this study comprises 728 research articles focused on the field of nanobiology.
The vast majority of these articles are written in English.
In total, the corpus contains 158,283 sentences and 3,762,791 tokens.
On average, each paper in the corpus consists of 217 sentences, and each sentence contains approximately 24 tokens.
This extensive dataset provides a rich foundation for in-depth analysis and research in the field of nanobiology.
The articles were first obtained in PDF format and the abstract and full text of each article was extracted using Grobid \citep{GROBID}.
Parts of the documents that are not considered as the core of the articles were excluded (e.g. References, Acknowledgment, Appendix).

\subsection{Ontology}
As resources, we used the NanoParticle Ontology for cancer nanotechnology research (NPO) \citep{NPO} and eNanoMapper (ENM) \citep{hastings2015enanomapper}, which are the two main ontology in the field of nanobiology.
As described in the ENM official documentation, ENM is an automatic extension of NPO and reuses several other ontologies including NPO, CHEMINF \citep{cheminfoonto}, CHEBI \citep{chebi} and ENVO \citep{ENVO}.
The NPO possesses 1904 classes and 81 properties, while ENM contains over 25k classes, 697 individuals and 55 properties (August 2023).
Since ENM is built automatically we used it as a secondary source to NPO, in order to minimize the risk of noise propagation.
The ontologies were used in CSV format, where each concept in the ontology had a unique key, definition, synonyms, and parent key.
These resources will be used for their subsumption relations and vocabularies, providing us with a taxonomy and lexical database.

\subsection{Labels and vocabulary}
To determine the labels our model will be trained to recognize, and their vocabulary, we used an iterative process of reducing the ontologies, expanding the obtained vocabulary and having every steps validated by domain experts.
Because of the large number of concepts in the NPO and ENM ontologies, the difficulty of finding a focus to start with and the fact that our aim is to create an automatically labeled corpus, we first retained only the concepts that presented at least one occurrence in our corpus (i.e. $\approx 30\%$ of NPO's and $\approx 10\%$ of ENM's).
Concepts that have never appeared in the corpus were discarded, and subsumption relations within the ontology were reconstructed to obtain a reduced ontology.

Using these reduced ontologies, three domain experts (cf. Acknowledgements) examined their structures and the remaining concepts.
Together, they identified five labels as being the core concepts of interest to the field of nanobiology, namely \texttt{Nanoparticle}, \texttt{Prop\-erty}, \texttt{Material}, \texttt{Event} and \texttt{Technology}.
Table \ref{tab:labelexample} presents a short description of each label, along with the core concepts combined under them (the number of their respective sub-concepts before expert selection is indicated between parenthesis) and a vocabulary extract in the last column.

The concepts corresponding to each label are taken as the root concept of ontology sub-trees.
We then amalgamated all the terms under the root concepts with all of the terms of all of its respective sub-concepts to built the labels vocabulary.
In any conflict between the NPO and ENM structure, NPO was prefered. 
The labels were subsequently subjected to a first detailed verification by the domain experts, who selected sub-concepts with relevant vocabulary only, which drastically downsized the number ().
In addition to verifying each label's vocabulary, they encountered six specific cases of terms under \texttt{Material} that they thought should be moved under \texttt{Nanoparticle}: \textit{buckyball}, \textit{carbon dot}, \textit{surface group}, \textit{dendrimer}, \textit{liposome} and \textit{fullerene}.

\begin{table*}[ht]
    \centering
    \begin{tabularx}{\textwidth}{lXXX}
        \hline
        \hline
        Label & Description & Core Concepts (\#sub-concepts) & Vocabulary extract \\
        \hline
        \texttt{Nanoparticle} & are physical structures, usually between 1 and 100nm in two or three dimensions, that present size related properties & Nanoparticle (68), Fiat Material (77) & \textit{nanocapsule, fluorescent carbon nanoparticles, carbon dot, nanowire}\\
        \hline
        \texttt{Property} & are physical and chemical functions that can be described using measurement & Realizable Entity (89), Application (102) & \textit{amphiphilic, hydrophilic, antioxidant, fluorescent}\\
        \hline
        \texttt{Material} & are the atoms and chemical compounds constituting nanoparticles and other studied objects & Chemical Entity (663), Material Entity (320) & \textit{thiol, gold, primary amine, carbohydrate}\\
        \hline
        \texttt{Event} & describes what is happening at a cellular level & Process (231) & \textit{mitosis, transcription, cell death, DNA modification}\\
        \hline
        \texttt{Technique} &  for preparing nanoparticles, measuring their characteristics and using them & Technique (63), Assay (171), Bioassay (37), Instrument (31), Application (102) & \textit{fluorescence spectroscopy, atomic force microscopy, gel electrophoresis} \\
        \hline
        \hline

    \end{tabularx}
    \caption{Description and examples of the chosen labels}
    \label{tab:labelexample}
\end{table*}

Table \ref{tab:label_Variant} presents the characteristics of the labels vocabulary.
Terms designates the vocabulary size for the label based on the ontologies lexicon.
Vocabulary indicates the size of the extended vocabulary based on terminological variations retrieval (cf. below), which includes the original terms.
Occurrences gives the raw frequency of all label's terms in our corpus.
Also, since this was obtained by reducing ontologies, the Depth and Width columns give an insight of the shape of each sub-tree.

\begin{table*}[ht]
    \centering
    \begin{tabular}{lrrrrr}
    \hline
    \hline
    Label & Terms & Vocabulary & Occurrences & Depth & Width \\
    \hline
    \hline
    \texttt{Nanoparticle} & 71 & 196 & 16,341 & 4 & 19 \\
    \texttt{Property} & 105 & 345 & 19,849 & 7 & 24 \\
    \texttt{Material} & 241 & 515& 74,688 & 11 & 36 \\
    \texttt{Event} & 56 & 210& 3,219 & 5 & 9 \\
    \texttt{Technique} & 65 & 172& 7,104& 7 & 19 \\
    \hline
    Total & 538 & 1,438 & 121,201 & & \\
    \hline
    \hline
    \end{tabular}
    \caption{Labels vocabulary sizes}
    \label{tab:label_Variant}
\end{table*}

After the expert determined the labels and corresponding terms, we recorded the variants of all the terms using FASTR \citep{FASTRjacquemin1997expansion}.
Given a list of terms and a corpus of texts, FASTR is able to extract the terminological variations using solely lexical, syntactical and meta-grammatical rules.
This tool is also able to account for variations in word order and part-of-speech changes.
It can deals with multi-word terms and is able to recognise variations in an expression (e.g. \textit{'molecular function'} $\rightarrow$ \textit{'functional roles of molecular'}).
Although the results of FASTR seemed rather accurate at first, a second round of expert validation of the vocabulary took place.
Out of 2,211 unique variations, experts reduced the number to 1,438 terms (i.e. 65\%) and thereby preserved the quality of the training data.

\subsection{Automatic corpus annotation}

We annotated the data for our distant supervision approach using Prodigy \citep{prodigy} under a research licence.
The annotation follows the CoNLL2003 \citep{sang2003introduction} standard, which uses the BIO annotation format.
The Occurrences column in Table \ref{tab:label_Variant} displays the number of annotation under each label in our corpus.

\section{Experimental methods}

The primary objective of our experiments is to test whether the model possesses good generalization capabilities, precision, and stability.
Therefore, we designed three distinct ablation studies to evaluate how dependent our approach is to the labels vocabulary.

\subsection{Exploring Existing Models}

To identify every entity in the articles, we first examined the results of the SciBERT model \citep{beltagy2019scibert}.
SciBERT is a widely pre-trained model for scientific articles, aiming at improving the expressivity of the model and save training time for downstream tasks.
We manually annotated 646 "naive" entities (i.e. "naive" meaning only distinguishing whether a span is an entity or not, not knowing which label the entity belongs to) related to the field of nanobiology in one article \citep{MA201613}, and then tried to use SciBERT for "naive" entity recognition on plain text.

The result is that SciBERT can identify almost all entities in the article.
Out of 646 entities related to the nanobiology field it can identify 638, which suggest a high recall capability (i.e. $\approx 0.99$ on the article tested).
However, SciBERT identifies a large number of entities that would be false positives in the field of nanobiology.
SciBERT identified a total of 2,976 entities, which gives 2,322 false positives that need to be filtered out suggesting a low precision value (i.e. $\approx 0.21$).
Examples of these false positives are : \textit{nanoscience}, \textit{construction}, \textit{convergence}, \textit{reduce}, \textit{Hayakawa}
A way to eliminated these false positives would be to match them with an ontology.
But this approach would lack several essential aspects: classification of entities into labels, possible confusion between concepts when trying to do so (e.g. in the ontology, \textit{dendrimer} is originally present under the concepts \texttt{Material} and \texttt{Nanoparticle}), coverage of the ontology vocabulary and so on.
Then, it does not eliminate the need for ontology reduction and expert involvement.

We also experimented with some existing models for BMNER in the Scispacy and Stanza libraries, but most of these are trained on specific corpora and entities, and perform poorly on NER tasks in the nanobiology field.

\subsection{Ablation experiment design}

In order to assess the dependency of the approach to the resource, as well as model generalization capabilities, we designed a set of ablation experiments.
As detailed in Table \ref{tab:label_Variant}, our five labels cover a list of 538 terms.
Each term has varying numbers of variants, ranging from 1 to over 10, resulting in a total of 1438 different terms.
In our ablation experiments, the terms were first randomly shuffled in each labels to minimize the risk of latent factors from affecting the experimental results, such as the terms being arranged in a specific pattern.
Then, the label's vocabularies are divided into five equal parts, noted as folds A, B, C, D and E.
Ablation of 33\% of the terms were also implemented with folds F, G and H.


\begin{table}[ht]
    \centering
    \begin{tabular}{llc|c}
    \hline
    \hline
    Fold & Abl.\% & Training set & Test set \\
    \hline
    A & 20\% & 126,834 & 21,427\\
    B & 20\% & 130,099 & 18,162\\
    C & 20\% & 136,708 & 11,553\\
    D & 20\% & 113,875 & 34,386\\
    E & 20\% & 129,580 & 18,681\\
    \hline
    F & 33\% & 109,259 & 39,002\\
    G & 33\% & 117,392 & 30,869\\
    H & 33\% & 120,103 & 28,158\\
    \hline
    \hline
    \end{tabular}
    \caption{Number of sentences in each Fold} 
    \label{Folds}
\end{table}

To create the training and test set for our ablation experiments, we selected the sentences based on the presence or absence of ablated entities.
This was done in order to ensure that the model would not confuse excluded entities for negative examples during the training, and to later test its capabilities to retrieve entities not encountered before.
As shown in Table \ref{Folds}, this resulted in training and test sets of various sizes, some being three time larger than others (e.g. test sets in folds D and C).
For instance, the training set in Fold A is composed of 126.834 sentences not containing a single ablated term, the test set is then made of the remaining sentences.
Thus, ablation experiments ensure that part of the test set consist in entities on which the model has not been trained. 
The approximate sentence ratio is 6:1 for 20\% ablation and 4:1 for 33\% ablation.

\section{Results and Analysis}

In this section, we first present the results of training the model on the full dataset, which performances aligned with our expectations.
Then, we detail the two random ablation experiments we designed, reducing the data by 20\% and 33\% respectively.
We noticed a significant fluctuation in the results of these experiments.
Therefore, we specifically analyzed the 20\% ablation experiment and based on these analyses, we proposed the hypothesis that including or excluding specific terms under our labels might have a significant impact on the precision and recall scores.
Following this, we designed a new frequency-based 10\% ablation experiment to explore this hypothesis.
The results from this new experiment successfully validated our conjectures.

\subsection{Training on the whole dataset}

\begin{table*}[ht]
    \centering
    \begin{tabular}{cccc|ccc}
    \hline
    \hline
    Epochs & Precision & Recall & F1 Score & New Entities & Correct Labeling & Precision \\
    \hline
    5.0 & 0.981 & 0.989 & 0.985 & 485 & 375  & 0.773 \\
    20.0 & 0.982 & 0.989 & 0.985 & 249 & 202  & 0.811 \\
    \hline
    \hline
    \end{tabular}
    \caption{Models trained using the entire dataset}
    \label{table:alldata}
\end{table*}

Initially, we trained the model on the complete dataset, carrying out a training with 5 and 20 epochs respectively.
Given the consistency of the training and validation datasets, the nature of this experiment is closer to a straightforward word-matching task.
Results are displayed in Table \ref{table:alldata}.
We found that the F1 score of the model reached a value 0.985, which is consistent with our expectations.

Subsequently, we performed a deep analysis of the model's generalization ability.
Our assumption was that it would be impossible to achieve 100\% coverage of terms in the corpus, so we had the model re-annotate the corpus.
Table \ref{table:alldata} thus also presents the number of unique new entities (i.e. ignoring the number of occurrences) identified by NanoNER, the number of correctly labeled entities and the associated precision.
Not cosidering the number of occurrences for these newly retrieved entities allows for a better estimation of the model generalization capabilities.
NanoNER achieves an precision value on new entities roughly around 0.8.
Additionally, we found that as the training epochs increased, the precision value on the newly found entities improved, but the number of new entities recognized decreased.
This indicates that the number of training epochs can be chosen according to the intended use, giving priority to recall or precision values on never-before-encountered entities.

\subsection{20\% Ablation experiments}

The primary objective of our ablation study is to further test the model generalization capabilities and dependency to the resource.
We employed early stopping for training, setting the number of epochs to 1 and the batch size to 32. 
The training results are presented in Table \ref{best -20}.
As we can see, the precision fluctuates around 0.79 and the gap between the highest and lowest precision values can be as high as 0.14.
The impact of the vocabulary ablation is even more visible on the recall: with an average score of 0.54, it has degraded considerably compared to the initial training.
This tend to indicates that the absence or presence of specific terms highly influence the quality of the trained model.
To address this, we conducted further exploration in Section \ref{5.4.1}.

\begin{table}[ht]
    \centering
    \begin{tabular}{cccccc}
    \hline
    \hline
    Fold & Precision & Recall & F1 Score  \\
    \hline
    A & 0.796 & 0.544 & 0.646 \\
    B & 0.826 & 0.575 & 0.678 \\
    C & 0.853 & 0.593 & 0.699 \\
    D & 0.756 & 0.424 & 0.544 \\
    E & 0.711 & 0.549 & 0.620 \\
    \hline
    Average & 0.788 & 0.537 & 0.637 \\
    \hline
    \hline
    \end{tabular}
    \caption{Models trained on 20\% ablation folds}
    \label{best -20}
\end{table}

\subsection{33\% Ablation experiments}

Next, we conducted experiments with a 33\% ablation of the terms.
The results in Table \ref{ablation 3} are as expected: the precision value remained around 0.8, but the recall rate dropped even further.
This is due to the higher number of terms excluded in 33\% ablation experiments compared to the 20\% ones.

\begin{table}[ht]
    \centering
    \begin{tabular}{cccccc}
    \hline
    \hline
    Fold & Precision & Recall & F1 Score \\
    \hline
    F & 0.752 & 0.332 & 0.461  \\
    G & 0.868 & 0.411 & 0.557  \\
    H & 0.847 & 0.460 & 0.596  \\
    \hline
    Average & 0.822 & 0.401 & 0.538 \\
    \hline
    \hline
    \end{tabular}
    \caption{Models trained on 33\% ablation folds}
    \label{ablation 3}
\end{table}

\subsection{Ablation experiments analyse}

Firstly, we conducted a generalization analysis on the ablation experiments from folds A to E.
We employed the same method as in the full data analysis: We initially listed all the deleted entities, and then had predictions made by the model on the entire corpus.
Subsequently, we matched all the entities predicted by the model with the deleted entities.
As depicted in Table \ref{Refound}, the model could stably rediscover up to 30\% of the deleted entities.
Considering that these results are not calculated based on occurrence rates, and it's quite challenging to discover the relationships between many obscure entities through deep learning, we believe these results are within our expectations.

\begin{table}[ht]
    \centering
    \begin{tabular}{cccc}
    \hline
    \hline
    Fold  & Retrieved & Ablated  & Recall  \\
    \hline
    A & 66 & 232 &  0.28 \\
    B & 89 & 353 &  0.25 \\
    C & 72 & 303 &  0.24 \\
    D & 77 & 249 &  0.31 \\
    E & 77 & 272 &  0.28 \\
    \hline
    Average & 76.2 & 281.8 & 0.27 \\
    \hline
    \hline
    \end{tabular}
    \caption{Refound}
    \label{Refound}
\end{table}

Next, we sought to analyze the variability between the folds.
We hypothesize that certain labels are excessively difficult, thereby affecting the overall performance of the task, we then evaluated each label separately.
Table \ref{tab:perflabels} displays the average recall and precision values of the labels over the different folds (detailled evaluation is available in Appendix \ref{sec:appendix}.
We found that \texttt{Nanoparticle} and \texttt{Event} have the most important variations in recall, while the variations in precision mostly concern \texttt{Nanoparticle} and \texttt{Technique}
Most of the average recall values over different labels are close to 0.54, but differences in average precision is more important when comparing the different labels.
\texttt{Material} and \textit{Property} displays scores over 0.90, \texttt{Nanoparticle} is around 0.83, but \texttt{Event} and \texttt{Technique} have significantly lower precision values (0.74 and 0.65 respectively).
These variations are analyzed as a result of the specific characteristics of the labels vocabularies.
E.g. \texttt{Event} and \texttt{Technique} contains terms from the scientific language that are not specific to the nanobiology field, and thus carry a higher risk of confusion.

\begin{table}[ht]
    \centering
    \begin{tabular}{l|cc|cc}
    \hline
    \hline
    \multirow{2}{3em}{Label} & \multicolumn{2}{c|}{Recall} & \multicolumn{2}{c}{Precision} \\
     & avg. & var. & avg. & var. \\
    \hline
    \texttt{Nanoparticle} & 0.57 & 0.23 & 0.83 & 0.13  \\
    \texttt{Material} & 0.55 & 0.13 & 0.92 & 0.06 \\
    \texttt{Event} & 0.53 & 0.22 & 0.75 & 0.10\\
    \texttt{Property} & 0.48 & 0.10 & 0.91 & 0.09\\
    \texttt{Technique} & 0.56 & 0.16 & 0.65 & 0.15\\
    \hline
    \hline
    \end{tabular}
    \caption{Average performances on each label}
    \label{tab:perflabels}
\end{table}

We also observed specific differences between the different folds for specific labels.
E.g. the recall for \texttt{Nanoparticle} is very high in fold A (i.e. 0.82), but significantly lower in fold E (i.e. O.19).
This suggests that certain words are highly important for specific labels.
In fact, in fold E, the term \textit{nanoparticle} was removed, which is not only a high-frequency term throughout the entire corpus, but also an essential word involved in different terms (e.g. \textit{nanoparticle}, \textit{gold nanoparticle}).

We believe that these high-frequency terms may have a great impact on the training of the model.
Therefore, removing these words during training might lead to a significant decline in the performance of the model.
To explore this hypothesis, we decided to conduct a third ablation experiment based on terms of frequency.

\subsubsection{10\% Ablation experiments}\label{5.4.1}


\begin{table*}[ht]
    \centering
    \begin{tabular}{lcccc}
    \hline
    \hline
    Training set & Precision & Recall & F1 Score & Corpus size \\
    \hline
    remove top 10\% & 0.599 & 0.279 & 0.381 & 21.2\% \\
    remove top 10\% + $mft^*$ & 0.668 & 0.371 & 0.430 & 51.0\% \\
    remove middle 10\%  & 0.802 & 0.683 &  0.738 & 99.0\% \\
    top 10\% only & 0.763 & 0.413 & 0.536 & 78.8\% \\
    \hline
    \hline
    \end{tabular}
    
    $^*mft$: most frequent term in each label
    \caption{Models trained on 10\% ablation}
    \label{10 ablation}
\end{table*}

We then sorted the terms according to their frequency in the corpus and conducted four more rounds of ablation, as shown in Table \ref{10 ablation}: in each label we tried removing 10\% of the most frequent terms (with one experiment reintroducing the first term in each label), removing the 10\% in the middle of the terms frequency and finally retaining only the 10\% most frequent terms in each label.
This approach limits the corpus exploitable as training and test sets, reflecting the distribution of the terms throughout the corpus.

Firstly, it appears that the frequency of the terms is a good indicator of the model dependency towards the annotation resource.
Indeed, precision and recall values are impacted proportionally to the rank of the terms removed.
Comparing the first and second rows also indicates that certain terms (i.e. the most frequent terms in each label) significantly impact the model's performance.
These terms may play a critical role in the classification task, or they could provide substantial contextual information, helping the model understand other related terms.
Regarding the third row's ablation experiment, although most of the corpus (99\%) was kept to train the model, the F1 score is far lower than 0.985 when trained with the full data.
This suggests that even terms with lower frequencies still significantly impact the model's performance.
These less frequent terms might carry specific information crucial for the model to understand and classify the text.
The fourth row's ablation experiment result indicates that retaining only the most common terms might lead the model to overly focus on these terms, overlooking other terms that may carry important information.
This could be because these common terms contain a lot of generic information but lack some specific, category-targeted information.


\section{Error analysis and improvement approaches}

\subsection{Sentence selection during ablation experiments}

In our ablation experiments, we sometimes encounter a scenario where a sentence contains an ablated term and an other one that is not, and thus we would like to remove only one of them.
In our experiments, we chose to exclude such cases to avoid having our model confuse ablated terms for negative examples.
But other strategies could be adopted to tackle this, such as the masking or replacement of tokens.

\subsection{Imbalanced dataset}

As observed during the ablation experiment, our corpus is highly imbalanced.
Some terms and labels appear more frequently than others.
Detailed evaluations of individual labels in ablation experiments are given in Appendix \ref{sec:appendix}.
This discrepancy might lead the model to over-learn from these high-frequency terms, thereby overlooking the importance of less frequent terms.
During future optimization, this problem could be tackled by using techniques such as oversampling or undersampling to balance the number of samples across different categories. 
In a study on NER using Wikipedia, \citep{al2015polyglot} adopted an approach that involves constructing a subset of the training corpus.
This strategy ensures that the conditional distribution of specific entity classes remains unaltered when they are positive examples, thereby significantly enhancing the model's performance across multiple languages.



\subsection{Vocabulary coverage in distant supervision}

Our training and evaluation assume that the ground truth annotations are accurate, which may not be the case in a distant supervision framework.
Thus there may be cases where the model is retrieving entities under the correct label, but that are considered false positives in our automatically annotated corpus.
We tried to reduce this effect by employing FASTR \cite{FASTRjacquemin1997expansion} to improve the coverage of our resource, but this required experts to filter out FASTR false positives.
Also, their is some known variations that FASTR is not able to retrieve (e.g. \textit{'iron oxide nanoparticle'} and \textit{'silicon dot'} are in \texttt{Nanoparticle} vocabulary, but their respective variations \textit{'iron nanoparticle'} and \textit{'silicon dot'} were not recognized). 

One solution would be to use a method known as knowledge distillation, which incrementally improves the model's performance through iterations, a training method used in the previously mentioned BOND \citep{liang2020bond} paper.
By using a teacher model to generate pseudo-labels for training the student model, the training effectiveness of the model is improved through repeated iterations.
Another solution is to manually annotate a sufficient  portion of high-quality data and then use it as validation set for the model.

\section{Conclusion}
In this work, we have introduced NanoNER, a tool for Named Entity Recognition in the field of nanobiology.
We designed an iterative process to determine the model labels and vocabulary using ontologies, domain experts and retrieving terminological variations.
This resulted in five labels, covering 1,438 terms, that allow for the automatic annotation of our corpus in a distant supervision approach.
Experiment analyses have demonstrated that our model can effectively identify entities of interest, both previously seen and new ones, in the field of nanobiology.
Given the complexity and abundance of technical terms in the field, our method shows promising applications in nanobiology.

We believe that this approach can be applied as is on other scientific fields, as it require only an ontology (or taxonomy) resource and minimum man-power.
This allow for the efficient training of NER models useful in downstream NLP tasks.

Ablation experiments showed a significant dependence of the model on the vocabulary used.
In future work, we could attempt data augmentation on the dataset to reduce its imbalance and enhance the model's training performance.
In addition, it is possible to use knowledge distillation for iterative model updates, which can reduce the false positive misjudgment during validation and improve the model's generalization capabilities. 


\section*{Acknowledgements}\label{sec:ackno}
We would especially like to thank the researchers who shared their expertise in nanobiology with us:
Raphaël Lévy (Professor of Physics at Paris Nord University), Nathanne Rost (Postdoctoral researcher in Nanobiosciences at Paris Nord University) and Federico Boem (Postdoctoral researcher in Philosophy of Sciences at University of Twente).

The \href{https://nanobubbles.hypotheses.org/}{NanoBubbles} project has received Synergy grant funding from the European Research Council (ERC), within the European Union’s Horizon 2020 program, grant agreement no. 951393.

\bibliography{custom}
\bibliographystyle{acl_natbib}

\clearpage
\onecolumn
\appendix
\section{Detailed ablation evaluation}\label{sec:appendix}
\begin{table}[ht]
\centering
\begin{tabular}{cc|ccc|ccc}
\hline
\hline
\textbf{Fold} & \textbf{Label} & \textbf{Positive} & \textbf{Total} & \textbf{Recall} & \textbf{Positive} & \textbf{Total} & \textbf{Precision} \\
\hline
A & nanoparticle & 2,557 & 3,108 & 0.822 & 4,645 & 4,821 & 0.963 \\
 & material & 8,952 & 16,022 & 0.557 & 15,433 & 15,821 & 0.975 \\
 & event & 233 & 342 & 0.68 & 398 & 444 & 0.896 \\
 & property & 2,999 & 8,287 & 0.363 & 4,033 & 4,241 & 0.951 \\
 & technique & 591 & 1,036 & 0.57 & 1,346 & 2,053 & 0.656 \\
\hline
B & nanoparticle & 1,466 & 3,309 & 0.443 & 2,854 & 4,082 & 0.699 \\
 & material & 8,959 & 14,566 & 0.615 & 14,961 & 15,377 & 0.973 \\
 & event & 203 & 419 & 0.484 & 341 & 428 & 0.797 \\
 & property & 2,531 & 4,122 & 0.613 & 3,964 & 4,101 & 0.967 \\
 & technique & 685 & 2,033 & 0.336 & 1,129 & 1,665 & 0.678 \\
\hline
C & nanoparticle & 1,346 & 1,900 & 0.708 & 2,475 & 2,590 & 0.956\\
 & material & 6,315 & 9,477 & 0.665 & 10,930 & 11,976 & 0.913\\
 & event & 409 & 1,095 & 0.373 & 554 & 843 & 0.657 \\
 & property & 1,385 & 3,410 & 0.406 & 2,229 & 2,295 & 0.971 \\
 & technique & 297 & 574 & 0.517 & 958 & 1,240 & 0.773 \\
\hline
D & nanoparticle & 2,371 & 3,364 & 0.705 & 4,369 & 4,934 & 0.885 \\
 & material & 8,876 & 29,268 & 0.301 & 12,920 & 14,016 & 0.922\\
 & event & 871 & 996 & 0.874 & 1,362 & 2,236 & 0.609 \\
 & property & 3,329 & 7,345 & 0.452 & 4,855 & 6,563 & 0.740 \\
 & technique & 1,156 & 1,361 & 0.849 & 2,784 & 3,547 & 0.785 \\
\hline
E & nanoparticle & 1308 & 6,826 & 0.192 & 2,652 & 4,064 & 0.653 \\
 & material & 8,237 & 13,339 & 0.617 & 12,670 & 15,670 & 0.809 \\
 & event & 249 & 996 & 0.251 & 366 & 477 & 0.767 \\
 & property & 1,749 & 3,056 & 0.573 & 2,921 & 3,251 & 0.898 \\
 & technique & 492 & 935 & 0.525 & 1,160 & 3,132 & 0.370 \\

\hline
\hline
\end{tabular}
 \caption{Ablation evaluation on individual labels}
 \label{tab:each label}
\end{table}

\end{document}